\documentclass[acmsmall,nonacm]{acmart}
\usepackage{todonotes}
\usepackage{multirow}

\AtBeginDocument{%
  \providecommand\BibTeX{{%
    \normalfont B\kern-0.5em{\scshape i\kern-0.25em b}\kern-0.8em\TeX}}}

\begin{document}

\title{Systemic Risk and Vulnerability Analysis of Multi-cloud Environments}


\author{Morgan Reece}
\affiliation{%
  \institution{Mississippi State University}
  \city{Mississippi State, MS}
  \country{USA}}
\email{mlr687@msstate.edu}

\author{Theodore Edward Lander, Jr.}
\affiliation{%
  \institution{Mississippi State University}
  \city{Mississippi State, MS}
  \country{USA}}
\email{tel127@cci.msstate.edu}

\author{Matthew Stoffolano}
\affiliation{%
  \institution{Rochester Institute of Technology}
  \city{Rochester, NY}
  \country{USA}
}\email{mds5648@rit.edu}

\author{Andy Sampson}
\affiliation{%
  \institution{National Security Agency}
  \country{USA}}
\email{agsamps@uwe.nsa.gov}

\author{Josiah Dykstra}
\affiliation{%
  \institution{National Security Agency}
  \country{USA}}
\email{josiah.dykstra@cyber.nsa.gov}

\author{Sudip Mittal}
\affiliation{%
  \institution{Mississippi State University}
  \city{Mississippi State, MS}
  \country{USA}}
\email{mittal@cse.msstate.edu}

\author{Nidhi Rastogi}
\affiliation{%
  \institution{Rochester Institute of Technology}
  \city{Rochester, NY}
  \country{USA}
}\email{nxrvse@rit.edu}

\renewcommand{\shortauthors}{Reece et al.}

\begin{abstract}
  \textbf{Abstract:} With the increasing use of multi-cloud environments, security professionals face challenges in configuration, management, and integration due to uneven security capabilities and features among providers. As a result, a fragmented approach toward security has been observed, leading to new attack vectors and potential vulnerabilities. Other research has focused on single-cloud platforms or specific applications of multi-cloud environments. Therefore, there is a need for a holistic security and vulnerability assessment and defense strategy that applies to multi-cloud platforms. We perform a risk and vulnerability analysis to identify attack vectors from software, hardware, and the network, as well as interoperability security issues in multi-cloud environments. Applying the STRIDE and DREAD threat modeling methods, we present an analysis of the ecosystem across six attack vectors: cloud architecture, APIs, authentication, automation, management differences, and cybersecurity legislation.
We quantitatively determine and rank the threats in multi-cloud environments and suggest mitigation strategies.
\end{abstract}

\keywords{Multi-cloud, Cybersecurity, Vulnerability Assessment, Risk Analysis}

\maketitle

\section{Introduction}

A multi-cloud environment is an end-user-created integration of systems that are hosted by two or more cloud providers. More than 75\% of modern enterprises are now using multi-cloud environments~\citep{checkpointsoftware}. The advantages of multi-cloud environments, including increased reliability and availability, drive this trend. With multiple cloud providers, organizations can select the best services and capabilities from different providers to meet their specific requirements. Multi-cloud environments can help to avoid vendor lock-in and provide resilience from downtime. Additionally, using multiple providers can help to distribute workloads and avoid service disruptions in the event of a failure or outage in one cloud provider.

Despite these advantages, security capabilities and features differ among the cloud providers making the environment more complex to integrate and manage. Different application programming interfaces (APIs), identity management, network security, automation, orchestration options, configurations, and logging technologies designed by various cloud platform providers compromise interoperability. This fragmentation in security capabilities has led to a highly compartmentalized approach to multi-cloud security, which can result in new attack vectors and potential vulnerabilities~\citep{duncan2020multi}. As a result, security professionals are often left with the challenge of creating their own custom solutions using unique cloud automation techniques or using third-party solutions to achieve interoperability. More cloud providers mean more individual compartmentalized security systems to configure, resulting in new attack vectors and potential vulnerabilities. These technologies were designed by various providers to ensure security but compromised on interoperability~\citep{durg_podder_2022}. The lack of interoperability is a root cause of new security issues in multi-cloud environments, as seen in the incident at Uber in September 2022. The attack compromised the company's multi-cloud setup by accessing an admin username and password stored in PowerShell scripts for automation. The same username and password gave the attacker access to all cloud services at Uber. This example underscores the importance of implementing robust security measures, including secure access management, and the need for an integrated approach to multi-cloud security that addresses the unique security capabilities and features of individual providers while ensuring interoperability across all providers~\citep{lovejoy_2022}.

Prior work on the security risk and vulnerability assessment for the multi-cloud environment is limited in scope and misses a holistic approach to analyzing and addressing security and vulnerability issues in a multi-cloud environment~\citep{overview_multicloud,CASOLA2018344,7019286}. In this paper, we perform a comprehensive and systematic study of the security risks on a multi-cloud platform and analyze various attack vectors that can be used to carry out an attack. We use the STRIDE and DREAD frameworks to study and quantify the risks posed by various attack vectors. Multiple attack vectors can be combined and used in tandem to create a more effective attack. For example, an attacker may use a phishing email to trick a user into revealing their login credentials and then use those credentials to gain access to a system. Once inside the system, the attacker can use other attack vectors, such as privilege escalation or malware installation, to take control of the system and carry out their objectives. By studying attack surfaces and the various attack vectors that can be used, we improve our understanding of potential threats and risks on multi-cloud platforms and develop more effective defenses and countermeasures to protect against them.

In this paper, we make the following contributions. Firstly, we conducted a literature review of cloud risk analysis frameworks. Secondly, we defined the multi-cloud architecture that was used for our research. Thirdly, we categorized the attack vectors in a multi-cloud environment within the STRIDE Threat Modeling Framework. Later, we calculated the quantitative risk for each attack vector to prioritize the risks associated with each attack vector prioritize the risks according to the DREAD Threat Modeling Framework. Finally, we identified risk mitigations such as multifactor authentication and privilege access management for each attack vector.

The rest of the paper is organized as follows. In Section~\ref{sec:background}, we present background material on  multi-cloud, risk analysis, and cloud vulnerability analysis. In Section~\ref{architecture}, we describe an exemplary multi-cloud application. We present attack vectors in Section~\ref{attackvectors} and vulnerability analysis in Section~\ref{sec:risk}. We summarize and conclude in Section~\ref{sec:conclusion}.

\section{Related Work \& Background}
\label{sec:background}

In this section, we cover relevant background research and related works on cloud threats, vulnerability analysis, risk analysis, and multi-cloud systems. We illustrate examples of multi-cloud environments utilized in healthcare, finance, and government.

\subsection{Cloud Computing Threats and Vulnerability Analysis}\hfill

Cloud computing platform delivers on-demand services to organizations over the Internet. Considered a cost-effective, scalable solution, they are the choice for many organizations that desire flexible access to information technology (IT) resources such as servers, storage, databases, and software applications. However, cloud computing also poses several security challenges, such as data breaches and identity theft, that must be addressed to ensure the confidentiality, integrity, and availability of cloud-based systems and data~\citep{chen2012data}. The National Institute of Standards and Technology (NIST) recommends several best practices for securing cloud-based systems and data, including strong access controls, data encryption, and continuous monitoring of the cloud environment to ensure confidentiality, integrity, and availability of the platform~\citep{badger2012cloud,mell2011nist}. However, organizations are increasingly adopting multi-cloud strategies and solutions for flexibility and portability to migrate, build, and optimize applications across multiple clouds and computing environments~\citep{google}.

Related works on cloud vulnerability analysis have focused heavily on single cloud providers or multiple independent providers. For instance, Torkura et al. proposed a technique for auditing and monitoring threats in multi-cloud using Amazon Web Service (AWS) and Google Cloud Platform (GCP)~\citep{torkura2021continuous}. While the solution supported data from both vendors, it did not address threats arising from connections between vendors. Other research examines the cloud components or features of a single provider or technology such as Docker and Kubernetes~\citep{zeng2023full}. Our work is focused on risks and vulnerabilities that arise from the interdependence and interconnection between cloud providers.

Current state-of-the-art does not provide a comprehensive analysis of the overall security risks, a thorough vulnerability assessment, or ways to address the challenges of implementing and maintaining security measures over time on multi-cloud environments. It focuses on specific aspects of multi-cloud security, such as data and application protection and designing security-focused applications. For instance, Hong et al. offered an approach for data protection and host and application protection as a Software-as-a-Service (SaaS) application in multi-cloud and federated cloud environments~\citep{overview_multicloud}. Casola et al. presented a fully automatable process for designing, developing, and deploying multi-cloud applications with a security-driven approach~\citep{CASOLA2018344}. Balasaraswathi and Manikandan proposed a cryptographic data splitting approach for securing information and storing metadata information in a private cloud to prevent unauthorized data retrieval~\citep{7019286}.

\subsection{Multi-Cloud in Practice}\hfill\label{practice}

Multi-cloud offers a wide range of application opportunities in multiple domains and use-cases. There is a considerable lack of previous research work on multi-cloud, multi-cloud security, etc. Before describing a multi-cloud architecture in Section~\ref{architecture}, we first present illustrative examples from healthcare, banking and finance, and government that showcase practical uses and security challenges of multi-cloud. 

\subsubsection{Healthcare Multi-Cloud Applications}\hfill

Multi-cloud environments are commonly utilized in healthcare cloud deployments~\citep{khattak2015security}. Many multi-cloud deployments develop organically through acquisitions. As healthcare organizations increasingly transition parts of their operations to cloud-hosted services, they may not move to the same hosting provider. Consequently, when hospitals merge or acquire other organizations with systems hosted in a cloud environment, they are likely with a different provider. At the same time, it is possible that different business units within an organization may decide to host their individual applications with another provider. These decisions may be driven by cost, performance features, and functions offered by the applications developers hosted in the cloud. An organization may even choose to host different portions of a software solution with another cloud provider. This architecture is reflected in Figure~\ref{fig:Multi-Cloud Blueprint}, and described in Section~\ref{architecture}.

Organizations need to share application information, which may introduce risk to the environment. The risk increases when the applications are hosted in different cloud providers because, as the information shared traverses the internet and is exposed to the public, where malicious actors can intercept it. Securing a multi-cloud architecture in the healthcare industry carries s higher risk than anticipated because of the value of the data stored, processed, and shared within the environment.

\subsubsection{Banking and Financial Multi-Cloud Applications} \hfill

Financial institutes have adopted cloud environments for various reasons and circumstances because of their mission type. However, despite this, there is always stress on the system's resilience across an economy since a mass failure can have a cascading negative impact for years. Maintaining this resilience has become a primary concern as cloud banking has developed to this point. For example, many smaller financial institutes, such as new credit unions, may use the same cloud service that specializes in storing bank data creating a singular point of failure. The primary solution to this concern has become a multi-cloud environment because it provides a way to diversify the impact of any singular crisis. Multi-cloud computing for banking has two components the onsite part and the offsite. The onsite cloud systems provide redundancy and store information regulators required storing in that area. The offsite environment offers availability and accessibility to a third party.

\subsubsection{Government and National Security Multi-Cloud Applications}\hfill

In the United States, cloud computing is a strategic imperative for government and defense. The Department of Defense Software Modernization Strategy states that ``The DoD Enterprise Cloud Environment is the foundation for software modernization. The multi-cloud, multi-vendor approach still holds true. The requirement for cloud across all classification domains, from enterprise to tactical
edge, is still valid''~\citep{DOD_Strategy_2022}. The Department of Veterans Affairs has also embraced multi-cloud, with plans to migrate nearly 350 of its 1,000 applications to cloud providers by 2024~\citep{GovMatters2021}. A combination of AWS and Microsoft Azure, together with on-premise data centers, provide infrastructure with governance, orchestration, and automation. In December 2022, the Department of Defence (DoD) awarded the Joint Warfighting Cloud Computing (JWCC) contracted to four cloud service providers: Microsoft, Amazon Web Services, Google, and Oracle~\citep{JWCC_awarded}. The awarding of the JWCC contracts to four providers enables the DoD to develop and deploy multi-cloud application service delivery environments.

\subsection{Risk Analysis}
Research in cloud infrastructure security and cloud-deployed applications continue to evolve and expand. Efforts to utilize standard analysis frameworks for assessing the security of these environments and systems are in place. The quantitative risk and impact assessment framework (QUIRC)~\citep{Saripalli2010QUIRC}, Optimized Infrastructure Services (OPTIMIS)~\citep{djemame2011risk}, and SEmi-quantitative BLO-driven Cloud Risk Assessment (SEBCRA)~\citep{Fitó2010} are all cloud risk assessment frameworks that aim to properly evaluate the security risk for a given cloud ecosystem to help prioritize risk mitigation efforts. 

Traditional risk analysis frameworks primarily focus on systems and applications. These frameworks are qualitative and use calculations based on limited data and, therefore, may produce skewed results that cause security leaders to prioritize environmental risk inaccurately. Akinrolabu et al. propose the Cloud Supply Chain Cyber Risk Assessment (CSCCRA) risk analysis framework evaluating multi-cloud environment~\citep{akinrolabu2019csccra}. The framework addresses inter-cloud communication and supply chain risk, examining code injection and manipulation that can compromise the integrity of the cloud deployment. CSCCRA also considers the dynamic nature of cloud environments and the limited visibility of the underlying infrastructure running virtualized systems. This framework addresses some of the shortcomings of risk management frameworks. However, gaps remain in assessing multi-cloud environments, namely in the lower system-level applications that are not SaaS-based. 

Afolaranmi et al. present various cloud service models (Infrastructure-as-a-Service (IaaS), Platform-as-a-Service (PaaS), SaaS, etc.) that discuss the current threat model and extend Open Web Application Security Project (OWASP), which focuses on risk analysis of software development~\citep{afolaranmi2018framework}. This paper also examines the challenges and benefits of multi-cloud while defining a framework for evaluating the risk involved in deploying applications in cloud environments. The focus on OWASP for identifying and assessing risk leaves a gap in securing inter-cloud communication within multi-cloud environments.

\subsection{Vulnerability Analysis Frameworks}\label{vuln_analysis}
Vulnerability analysis seeks to identify specific areas of attack on systems and applications and the mitigations. The uniqueness of vulnerabilities in cloud environments demands specialized approaches, which many proposed frameworks, processes, and procedures aim to address. Recent work in cloud vulnerability analysis primarily focuses on Infrastructure-as-a-Service (IaaS) offerings. For instance, Torkura et al. monitored the cloud infrastructure and enumerated the known vulnerabilities in cloud-specific and cloud-aware applications~\citep{torkura2020cloudstrike}. This approach consolidated the vulnerabilities of the infrastructure and the applications where an individual could prioritize the identified vulnerabilities. Ristov et al. proposed a framework to identify known vulnerabilities published in a public database. It reviewed dashboard and virtual machine known vulnerabilities, which may or may not have been patched by the software vendors ~\citep{ristov2014security}. It also explored the ability for cross-tenant communication with the need for improved application management. Kamongi et al. proposed VULCAN, a vulnerability assessment framework in ~\citep{kamongi2013vulcan}. The framework enables security professionals to create and deploy exploits targeted to expose vulnerabilities in cloud systems. VULCAN leverages an ontological knowledge base, natural language processor, Ontology Vulnerability Database, and vulnerability class index. A significant feature of VULCAN is its integration with exploitation tools to enable active vulnerability testing of possible vulnerabilities.

Research on cloud environments is extensive, yet it mainly focuses on single-cloud provider environments and disregards multi-cloud environments. Even though risk analysis focuses on single-cloud environments, industries such as finance, healthcare, and governments are transitioning towards multi-cloud environments. In the following section, we describe the multi-cloud architecture and blueprint that we used for our research and analysis.

\section{Multi-Cloud Architecture and Blueprint}\label{architecture}

\begin{figure} [ht]
    \centering
    \includegraphics[scale=0.5]{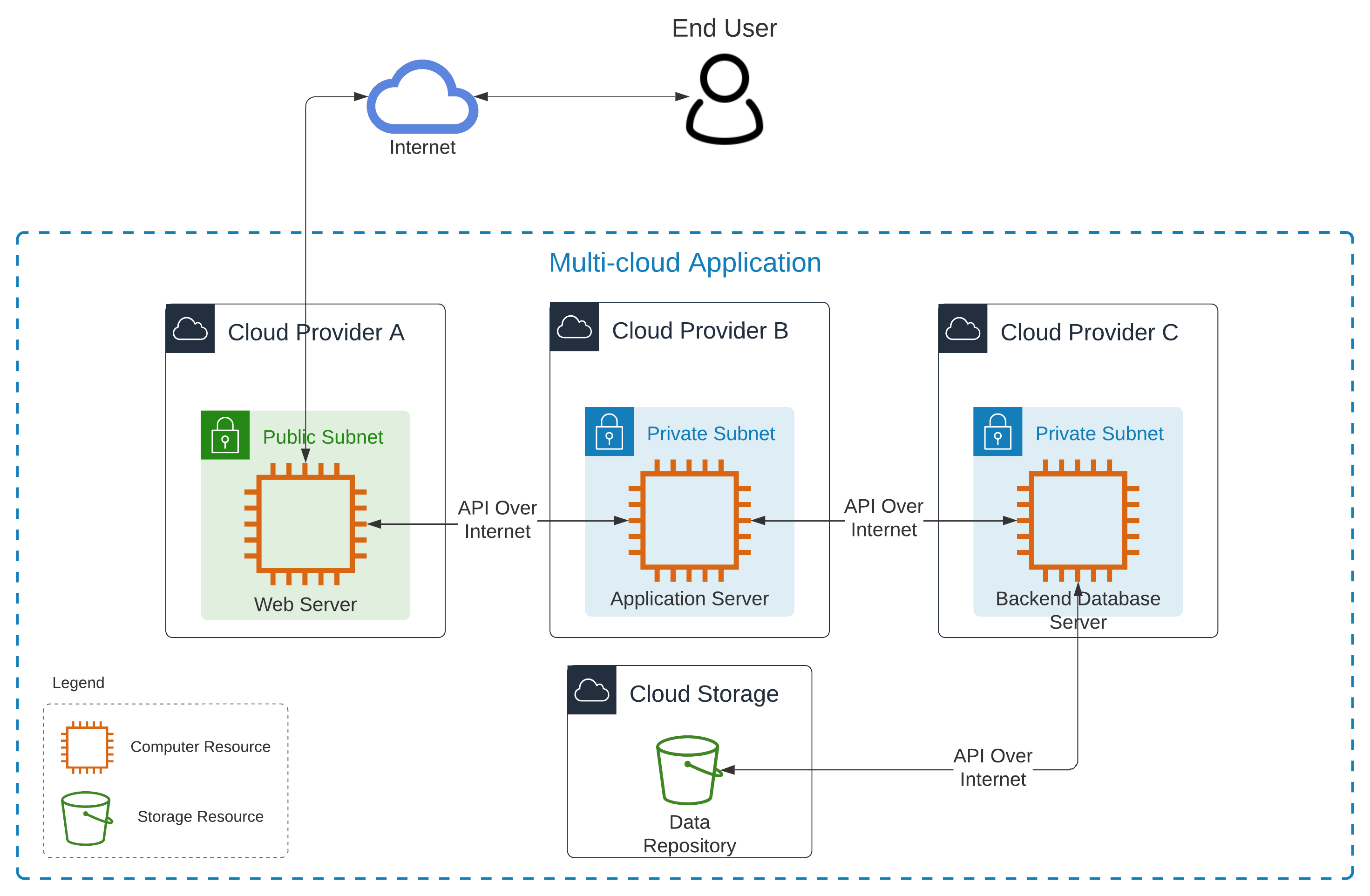}
    \caption[A typical multi-cloud architecture and application.]{\textbf{Three-Tier Web Application Architecture.} A multi-cloud architecture to support distributed functional services across cloud providers with Web Server as a public-facing, Application Server executing business logic, and Database Server managing data.}
    \label{fig:Multi-Cloud Blueprint}
\end{figure}

The design of a multi-cloud architecture should consider the possibility of dispersed locations for the distinct modules and services of the multi-cloud system. The benefits of utilizing a 3-tier architecture are~\citep{kambalyal20103}:
\begin{itemize}
    \item able to update each layer individually
    \item modular construction means that each layer is replaceable
    \item superior performance in high volume environments
    \item highly scalable 
    \item business rules are easier to implement in the application server
\end{itemize}
In this section, we outline a 3-tier~\citep{IBM2023} multi-cloud web application architecture. This architecture is derived from the study of multiple multi-cloud web applications described in Section~\ref{practice}. In Section~\ref{sec:risk}, we use this architecture to conduct risk analysis and create an attack vector testbed environment for our research investigation. 

These modules expand the overall attack surface of the entire system. Therefore, minimizing the attack surface when designing, configuring, and deploying multi-cloud systems is critical. This architecture has been showcased in Figure~\ref{fig:Multi-Cloud Blueprint}.

The partitioning of the 3-tier architecture across multiple cloud providers allows for each server to be hosted by a different cloud provider than the other servers. The different servers of the architecture are:
\begin{itemize}
    \item Web Server
    \item Application Server
    \item Database Server
    \item Data Repository
\end{itemize}

In the architecture, the Web Server manages the user interaction with the application. The Application Server processes the business logic of the system. The Database Server executes the data queries received from the Application Server. Finally, the Data Repository stores the system data from manipulation and retrieval. In the following sections we describe each of these servers in depth and their operations within the system.

\subsection{Web Server}\hfill

The Web Server hosts the public interface where users can interact with the application service. Usually, it delivers web pages to the user through a browser on a public subnet where users can access the service or the site. The software running on this server is designed to be accessed by privileged and non-privileged users and, therefore, presents a set of challenges. Each user type can access different levels of classified web pages and data. Because of the inherent expectation that anyone, including an attacker, can communicate with this server, the security configuration is critical. This server collects information from users who access the front-end service and transmits it to the application server through an application programming interface (API). The Application server then responds with the relevant data for the front-end server to present to the user. 

\subsection{Application Server}\hfill

The application server is located in a private subnet, where it receives requests from the web server over a secure API. Once the request is received, the application server processes them through applications and services and may send subsequent requests to the database server over an API. Once the application server has finished processing the requests and received the requested data from the database server, the application server sends the data and results to the front-end server. 

\subsection{Data Layer}\hfill

The third tier of our architecture is the Data tier. As a whole, this tier provides data access and management services to the application tier.  In our architecture, we have broken this tier is in two parts: 

\begin{itemize}
    \item 3a. Database Server 
    \item 3b. Data Repository
\end{itemize}

\subsubsection{Database Server}\hfill

The database server is also protected within a private subnet, which should only be accessible from an application server over an API. It carries out all operations of adding, modifying, and removing data from the database and executing queries on the database to retrieve requested data from the application server. However, the actual data repository can be local to the database server. Nonetheless, many a time, it is stored in a large cloud storage service system that may reside in a different cloud infrastructure. The database server employs an API to communicate with the data repository.

Isolation of the database servers in a micro-segmented network allows for fine-grain control of what systems can communicate with the database server and the protocol used in the communication. With the database server isolated in the private subnet, the storage for the database can be served by another cloud provider. The storage must also be tightly controlled, allowing specific systems and users to access the storage. Because different cloud providers can host the database server and storage, the communication will be exposed to the public internet. The data between the database server and storage must be encrypted, protecting the data in transit.

\subsubsection{Data Repository}\hfill

Data repository for the database is typically stored within a specialized cloud storage service optimized to provide fast and efficient data operations. The warehouse receives data and operational commands over an API. The repository is secured from inappropriate access from unauthorized users. Limiting access to the repository protects the confidentiality and integrity of the data.

In the next section, we will discuss the overall methodology that was used to conduct the modeling and analysis of the attack vectors.

\section{Risk \& Vulnerability Analysis Research Methodology}\label{methodology}

Security problems can arise when organizations integrate multiple cloud platforms due to exposure to attack vectors. To manage the vulnerabilities and implement mitigations, organizations should first establish a fundamental understanding of the relevant concepts, set up a multi-cloud test-bed environment, identify potential attack surfaces, and conduct a formal risk and vulnerability analysis. Implementing the resulting mitigations can create a more secure environment, ultimately improving the overall performance of the multi-cloud environment.

A \textit{two-pronged} methodology was used in the investigation and analysis of the attack vectors (Section~\ref{attackvectors}) and the risk associated with the attacks (Section~\ref{sec:risk}). First, using the STRIDE framework guided us through the \textit{identification and categorization of the attack vectors}. The STRIDE framework also supported and enabled the identification of strategies to help mitigate the risks associated with each attack vector. The second prong of our analysis strategy was using the DREAD threat modeling framework to conduct a quantitative analysis of the risk associated with the attack vectors. The results of the DREAD threat modeling framework produced risk values that showed how 'risky' the attack is to a multi-cloud implementation of a web application.
We will next present the STRIDE and DREAD frameworks and how they are used in threat modeling.

\subsection{STRIDE}\hfill
\label{STRIDE}

STRIDE is a framework used for modeling security threats to systems, networks, and infrastructure~\citep{hernan2006threat}. Even though it was initially developed to be used in the development life-cycle, it has become a mature framework for modeling threats at every level in a digital environment~\cite{SEI-blog}. STRIDE is an acronym for the six categories of threats that present challenges to the security of various types of digital systems. These categories guide the identification and mitigation of threats in the environment being evaluated. 

\begin{itemize}
    \item \textit{Spoofing identity} : pretending to be someone you are not
    \item \textit{Tampering with data} : unauthorized changing data within a system
    \item \textit{Repudiation} : disputing the authenticity of a communication
    \item \textit{Information disclosure} : revealing information that is confidential
    \item \textit{Denial of service} : preventing the normal execution of a system
    \item \textit{Elevation of privilege} : gaining more accesses than authorized
\end{itemize}

STRIDE was selected as one of the threat modeling frameworks for this research because of the extensive experience the team has with STRIDE as well as the wealth of supporting information and associated tools that could be utilized in applying the framework to the multi-cloud architecture. 

\subsection{DREAD}\hfill
\label{DREAD}

The DREAD Threat Modeling framework, like STRIDE, was also developed by Microsoft, but instead of categorization and mitigation of threats, DREAD is used to perform quantitative analysis on the identified threats. DREAD is an acronym for the six categories that each threat is scored on~\citep{DREAD_risk}.

\begin{itemize}
    \item \textit{Damage} : How Much Damage Could the Attack Cause?
    \item \textit{Repeatability} : How Easily Can the Attack Be Reproduced?
    \item \textit{Exploitability} : What’s Required to Launch the Attack?
    \item \textit{Affected Users} : How Many People Would the Attack Affect?
    \item \textit{Discoverability} : How Easy Is the Vulnerability to Discover?
\end{itemize}

The scores range from 0-10 depending on how relative value of the answer to the question associated with the category in the list above. After the threat has been scored, the values are then added together to give a total risk score. 
\begin{itemize}
    \item Critical (40-50): Critical risk; address immediately
    \item High (25-39): Severe risk; consider for review and resolution soon
    \item Medium (11-24): Moderate risk; review after addressing severe and critical risks
    \item Low (0-10): Low risk to infrastructure and data
\end{itemize}
The risk score can then be used to prioritize the project to implement the mitigation for the particular threat that was identified in the STRIDE threat modeling effort.
In the next section, the results of the identification and categorization of attack vectors through the execution of the STRIDE threat modeling.

\section{Identification \& Categorization of the Attack Vectors}\label{attackvectors} 

\begin{figure} [!ht]
    \centering
    \includegraphics[scale=0.5]{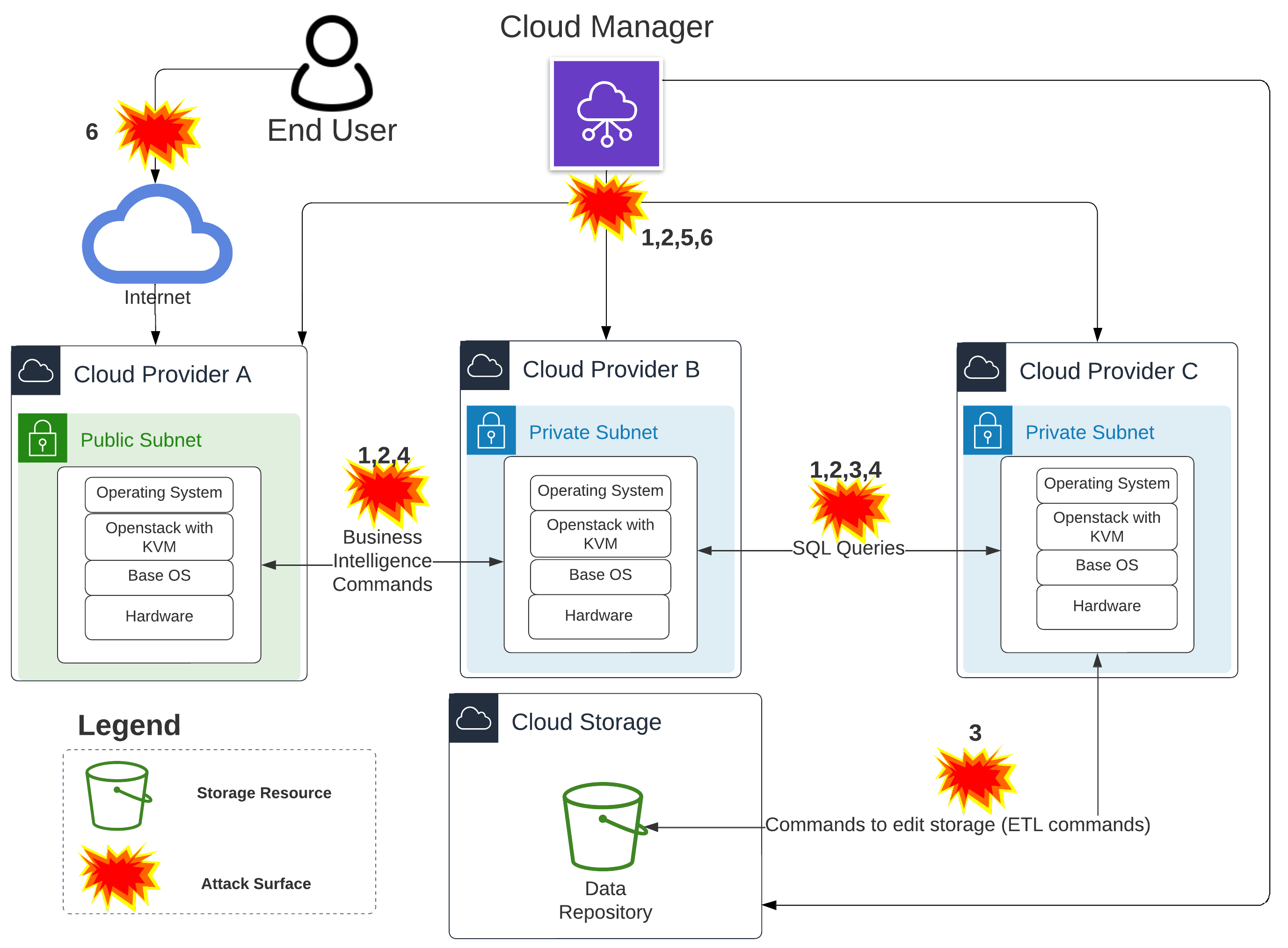}
    \caption[Multi-Cloud Attack Vectors Blueprint.]{\textbf{Multi-Cloud Attack Vectors Blueprint.} It demonstrates potential attack surfaces and shows the more specific inner workings of each part of the multi-cloud. The bang represents where a single attack vector or a set of attack vectors could occur. The numbers near the bang symbols represent the different attack vectors. (1:architecture, 2:API, 3:authentication, 4:automation, 5:SLA, 6:regulatory).}
    \label{fig:Mutli-Cloud Attack Vectors}
\end{figure}

In this research, we analyze attack surfaces in a multi-cloud environment and the various attack vectors that comprise them. Figure~\ref{fig:Mutli-Cloud Attack Vectors} shows examples of attack surfaces and vectors that could occur in a multi-cloud environment. From Section~\ref{architecture}, we understand that a multi-cloud environment can have numerous attack surfaces, allowing for many methods of attacks. The attack vectors considered in this research are:

\begin{enumerate}
    \item Architecture
    \item Application Programming Interfaces (APIs)
    \item Authentication
    \item Multi-cloud automation and orchestration
    \item Service level management differences
    \item Cybersecurity legislation
\end{enumerate}
\hfill\\
This implies that combining these attack vectors creates an entire attack surface to target the multi-cloud network. While Figure~\ref{fig:Mutli-Cloud Attack Vectors} does not show a complete analysis of the various attacks that could occur in a multi-cloud, it highlights the importance of the different attack vectors we have addressed in our research.

\begin{figure} [!ht]
    \centering
    \includegraphics[scale=0.5]{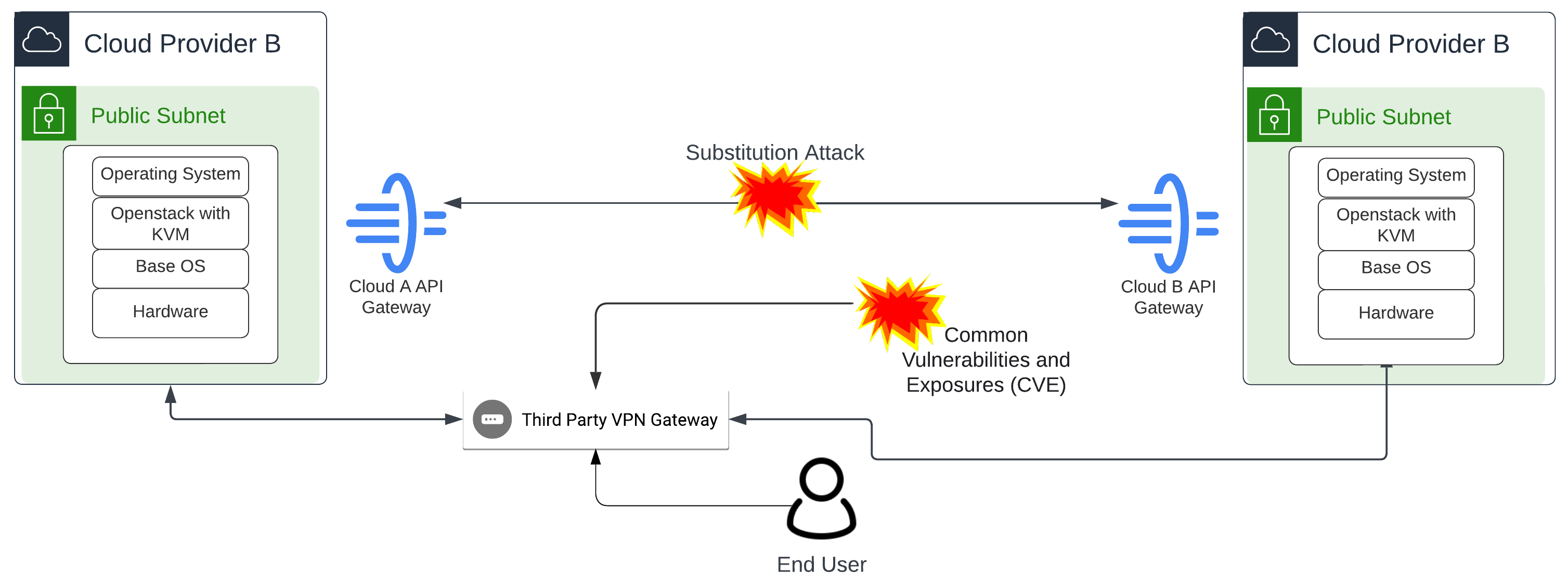}
    \caption[Architectural attack vectors in multi-cloud]{\textbf{Architectural attack vectors in multi-cloud.} Attack vectors targeting architectural vulnerabilities in multi-cloud environments seek to exploit weaknesses in design due to differences in cloud implementations and security protocols.}
    \label{fig:Different Architectures.}
\end{figure}

\subsection{Architectural Vulnerabilities in Multi-Cloud} 

In a multi-cloud environment, differences in implementing fundamental computing paradigms and standards can cause significant security issues and introduce potential attack vectors~\citep{Jahan2017}. For instance, encryption offerings and capabilities of different cloud platforms make it challenging for an organization to standardize a single security mechanism for storing and managing data. In another instance, most cloud platforms provide default defense against Denial of Service (DoS) but have difficulty measuring multi-cloud implementation's effectiveness~\citep{Mcloud_effective}.

Several types of network traffic need examination. One way to connect to cloud environments is through a Virtual Private Network (VPN). Some cloud providers provide VPN capabilities, but many allow for third-party VPNs. This flexibility enables an organization to access the whole cloud ecosystem using one VPN, making it vulnerable to common VPN server vulnerabilities. In the event of a VPN infiltration, various attacks become possible such as a substitution attack. This attack is primarily used when the key is not changed, and the attacker understands the communication enough to know what they must edit in the encryption to cause sufficient damage~\citep{kang2009vulnerabilities}.

The ever-expanding list of Common Vulnerabilities and Exposures (CVEs) is paramount to this research. These CVE vulnerability targets include the guest operating system (OS), hypervisor, and host OS~\citep{bazargan2012state}. Each includes its own vulnerabilities and problems; combining them creates a quickly expanding list of potential problems and issues to fix. For example, hypervisors could be at risk of several different types of attacks, one of which includes denial of service. By gaining access to the hypervisor, an attacker would be able to access data transported to and from the cloud provider and be able to remotely execute commands~\citep{sheinidashtegol2017performance}. Exacerbating the risk of having unpatched vulnerabilities is having known vulnerabilities across a multi-cloud environment. 

\begin{figure} [!htp]
    \centering
    \includegraphics[scale=0.5]{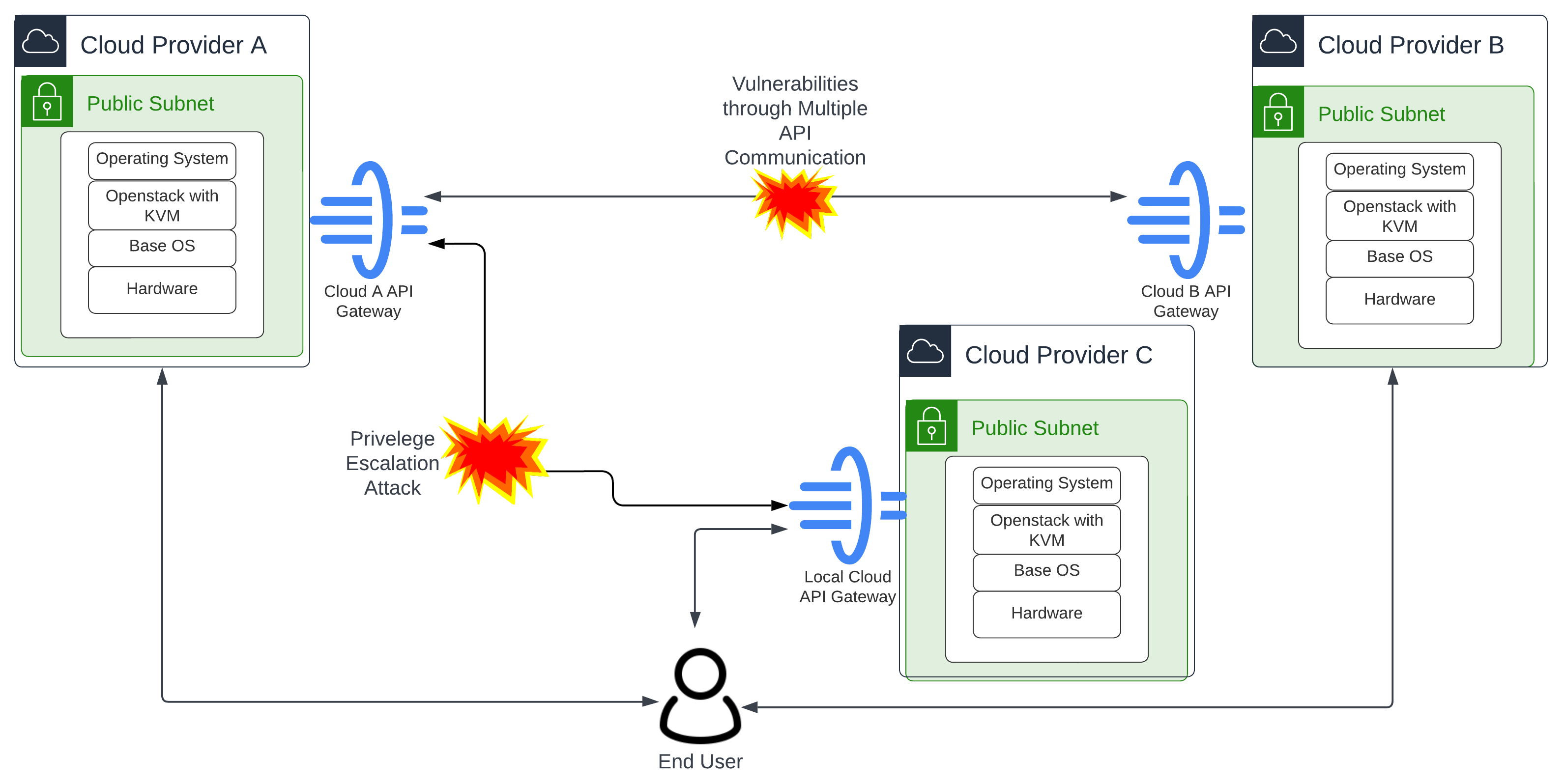}
    \caption[API Attack Vector]{\textbf{API Attack Vector:} Attacks against APIs seek to interrupt, disrupt, corrupt, and seize data being transferred to an application.}
    \label{fig:APIAttackSurface}
\end{figure}

\subsection{Application Programming Interface (API) vulnerabilities}\label{API-Threat} 

The fundamental basis of a multi-cloud architecture is the ability to build applications that can leverage the advantages of disparate cloud service providers. These applications are dependent on the interoperability provided by cloud APIs. 

A multi-cloud application that uses two different providers running their own independently designed API may have conflicts making it vulnerable to attacks causing interruptions of service. Another example is using a malformed packet, meaning that crafting the packets in a specific way may enable an attack. Privilege escalation can also be exploited through APIs in which two clouds distribute user privileges and look for ways for someone to elevate their privileges on a different cloud. The attacker also can send malformed packets to the API to corrupt, interrupt, or override the normal operation of the application~\citep{ariffin2020api}.

\begin{figure} [!htp]
    \centering
    \includegraphics[scale=0.5]{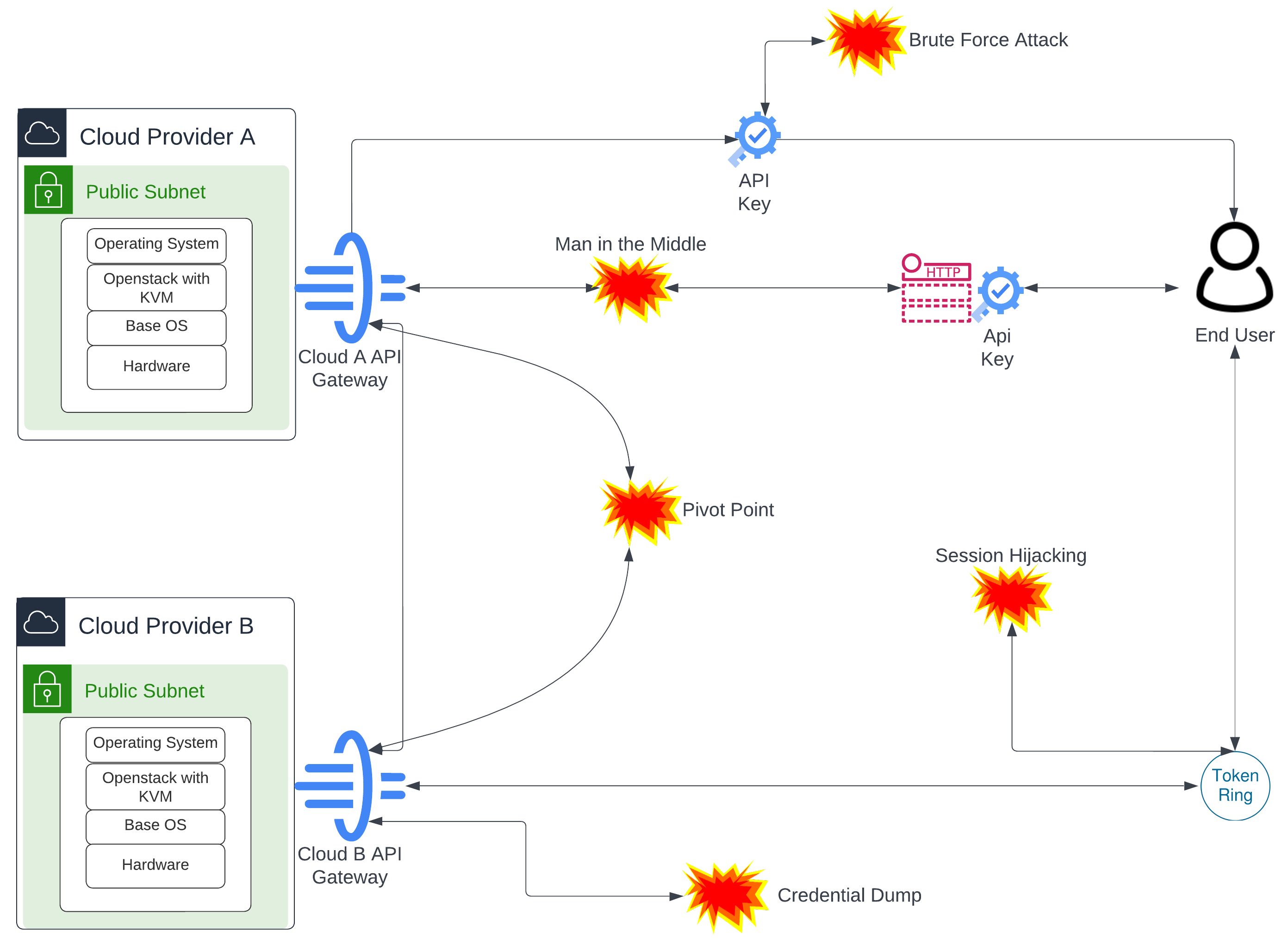}
    \caption[Authentication Attack Vector]{\textbf{Authentication Attack Vector:} Identification of attacks that could occur through an authentication attack vector.}
    \label{fig:Authentication Attack Vector}
\end{figure}

\subsection{Authentication} 
The user and application level authentication implementations can differ significantly from one cloud offering to another~\citep{choksi2014comparative}. Incorrect implementation can significantly increase risk and introduce vulnerabilities~\citep{dalton2009nemesis}. For example, uniform Identity and Access Management (IAM) in multi-cloud environments are challenging, as different cloud platforms define and set identity independently~\citep{INDU2018574}. Uniformly securing multi-cloud resources by applying a set of Access Control Lists (ACLs) is difficult as each cloud provider's security implementation differs. A recent trend in the security industry is the thought that ``identity is the new perimeter''~\citep{edwards2013identity}. The increase of remote and mobile workers has elevated the priority of identity and end-user devices in defining a security strategy. Verizon's annual Data Breach Investigations Report states that nearly 50\% of breaches involve stolen credentials or brute force attacks highlighting the risk associated with this attack vector~\citep{DBIR2022}.

\begin{figure} [!ht]
    \centering
    \includegraphics[scale=0.6]{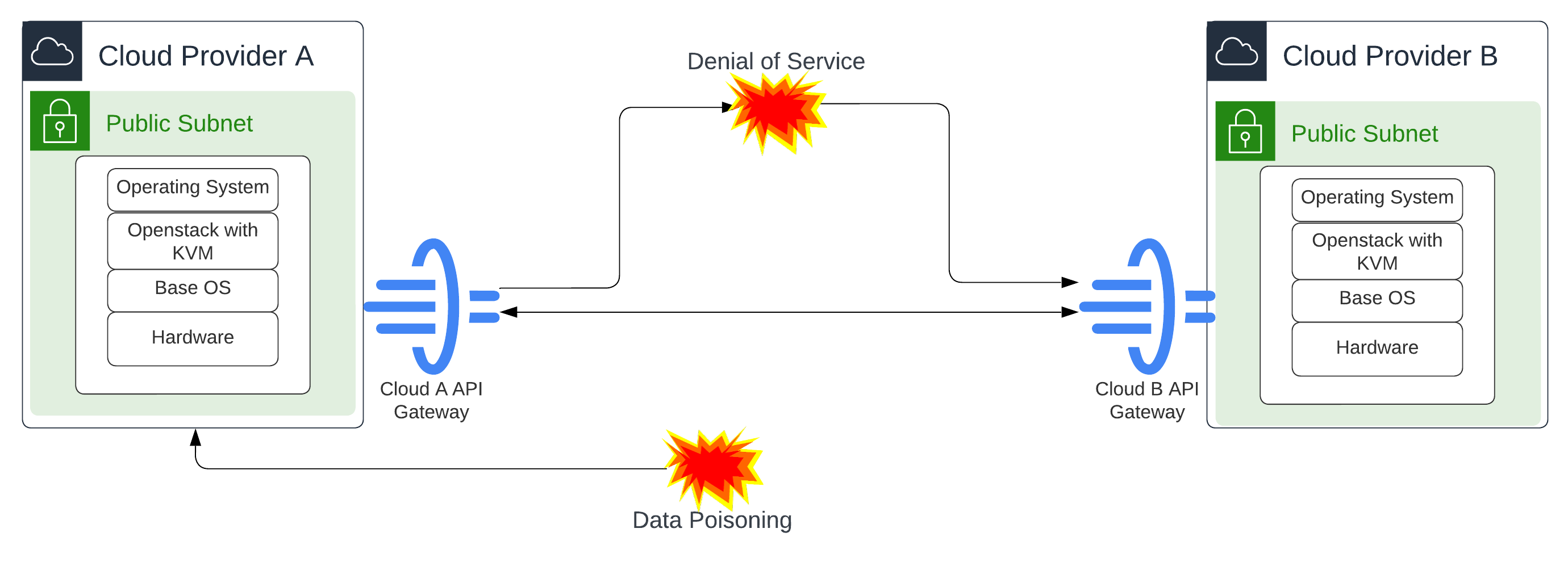}
    \caption[Automation Attack Vector]{\textbf{Automation Attack Vector:} Illustration of different attacks that could occur through automation.}
    \label{fig:Automation Attack Vector}
\end{figure}

\subsection{Multi-Cloud Automation and Orchestration}
Many organizations utilize and depend on cloud automation and orchestration for system optimization. Studying interactions of automation and the systems they are controlling is necessary to fully identify potential emergent security issues, mitigate those risks, and protect their valuable assets~\citep{RAMACHANDRA2017465}. Paladi discusses the security landscape of orchestration in multi-cloud deployments~\citep{Paladi2018}. The threat model presented in Paladi defines different layers in the infrastructure where attacks are targeted. Identified in their research, the attack vectors that are unique to automation in multi-cloud environments are attacks on the hardware abstraction layer and the software-defined infrastructure. The object of most attack vectors on the automation in multi-cloud environments is to interrupt, limit, or control the coordinated execution of the automation systems to optimize the execution of the multi-cloud environment.

\subsection{Difference in Management Schemes}\label{mgmt-schm}
\hfill\\
\begin{figure} [ht]
    \centering
    \includegraphics[scale=0.5]{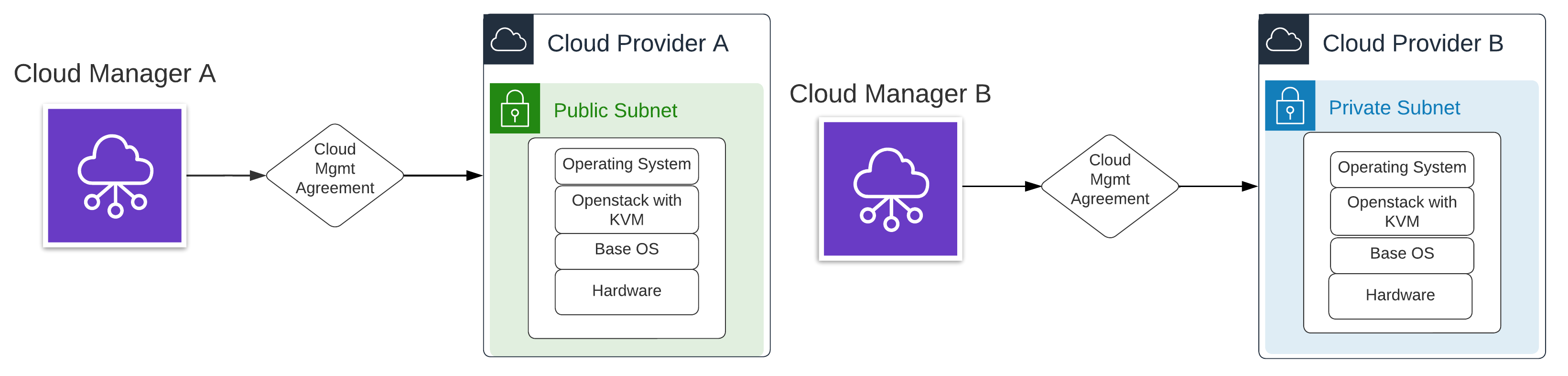}
    \caption[Difference in Management]{\textbf{Difference in Management:} When targeting management differences, attackers can leverage different multi-cloud systems to gain full access to an entire system by causing service outages and other attack methods listed above.}
    \label{fig:Difference in Management}
\end{figure}

The examination of the difference in management schemes between cloud service providers involves a contrasting study of Service Level Agreements (SLAs), systems operational, monetization features, and dynamic scaling features~\citep{forell2011cloud}. The difference in management can negatively affect a cloud environment and can be utilized as an attack vector when targeting a multi-cloud system. 
Cloud management agreements allow for confirmation between a cloud provider and a customer that some level of service will be provided. The service should include a form of monetization, auto-scaling, and a Service Level Agreement~\citep{forell2011cloud}. Figure~\ref{fig:Difference in Management} illustrates the relationship between two different cloud providers. Each cloud provider has its own cloud management agreement that supports its infrastructure and secures it. When utilizing more than one cloud provider, each Cloud Management Agreement needs to coincide. For example, cloud provider A in the figure is under a denial of service attack. This would affect the auto-scaling of the cloud system as well as create a money issue for the cloud provider and also make the service unavailable. If cloud provider B is receiving information from cloud provider A, cloud provider B is unable to receive the information and cannot complete its processes~\citep{alzain2012cloud}. In addition, when you have different management schemes, this will also create a multitude of different hypervisors, operating systems, and hardware being used by each cloud provider. This could also be a method of exploit for attackers looking to attack through a difference in management.

\begin{figure} [h]
    \centering
    \includegraphics[scale=0.48]{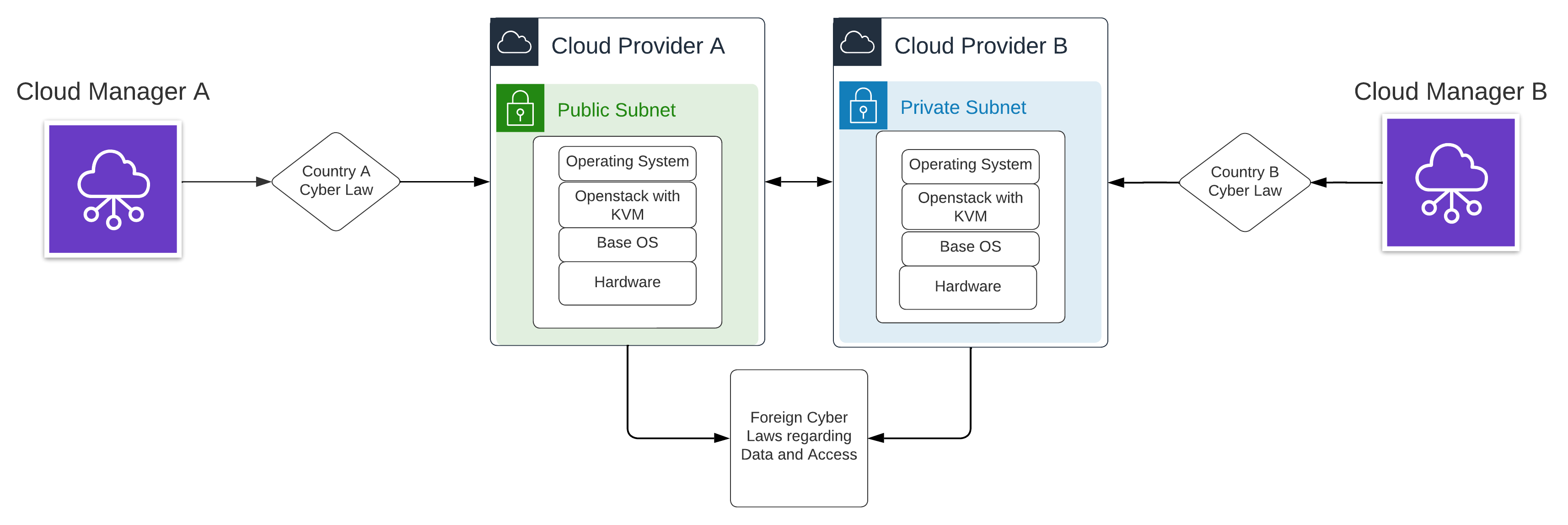}
    \caption[Different Cybersecurity Legislation]{\textbf{Different Cybersecurity Legislation: } When viewing legislation, attack vectors are created and grow exponentially as you increase the number of countries involved in a multi-cloud environment. When a regulation is missed, there is a severe financial threat to businesses.}
    \label{fig:Different Cybersecurity Legislation}
\end{figure}

\subsection{Mismatch of Cybersecurity Legislation} \label{mismatch}\hfill\\
Different cybersecurity legislation could cause many issues for multi-cloud. Differences in geographical location, compliance obligations, and jurisdictions (specifically data sovereignty laws) can potentially increase risk in multi-cloud environments. Implications of data privacy and control are paramount when looking at the different legal environments and laws. In addition, different server components and multi-cloud applications are in different locations worldwide. This could create many different issues. Having a clear agreement on foreign regulations between two countries will allow for the use of the multi-cloud environments, as illustrated in Figure~\ref{fig:Different Cybersecurity Legislation}. However, if country A has high standards and procedures for the collection of information from online sources and country B has strict laws on the availability of data to specific organizations, both countries would have to be in agreement that all required laws are met. If one of the countries refuses to cooperate, then the two different cloud providers cannot communicate information~\citep{rios2019service}. Different data privacy laws in Europe and the United States create friction. Many European countries are under the General Data Protection Regulation (GDPR)~\citep{europeanunion2020,tankard2016gdpr}. European citizens under the GDPR have the ``right to be forgotten,'' and organizations have an obligation to abide by the request of an individual not to keep a hold of data~\citep{politou2018backups}. Companies have to meet the standards stated in the GDPR as well as the laws and regulations of other countries that have ties to the multi-cloud. This could create many different foreign policy issues as well by not abiding by a specific country's laws that are set for its citizens~\citep{rios2019service}.

\section{Risk \& Vulnerability Analysis}
\label{sec:risk}

In this section, we will present the multi-cloud testbed used in our research and analysis, highlighting a specific use case that was used in our analysis. We also discuss the risk analysis methodology, results, and mitigations for each identified attack vector.

\subsection{Multi-Cloud Security Testbed}
\begin{figure} [!ht]
    \centering
    \includegraphics[scale=0.13]{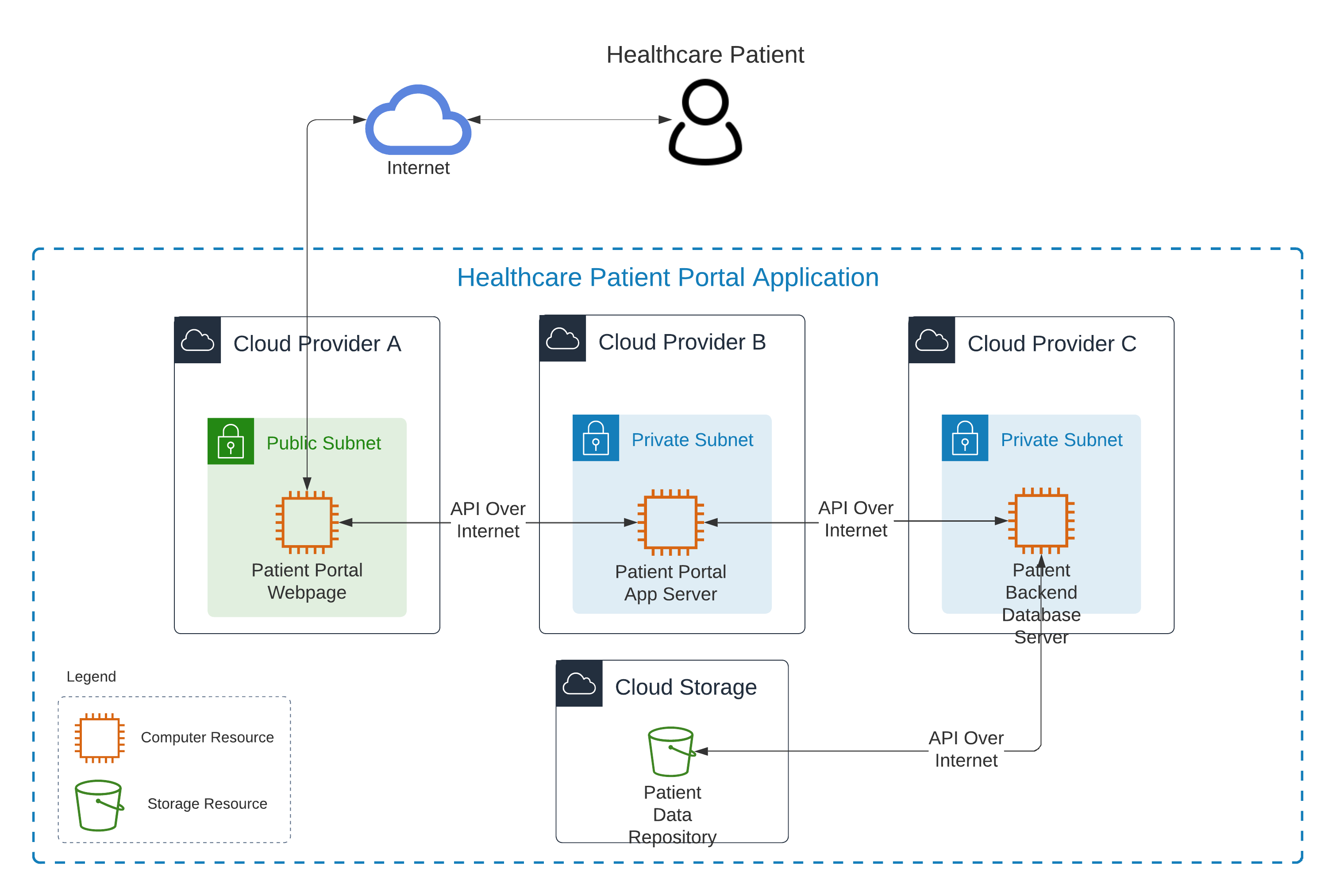}
    \caption[Healthcare App Deployment]{\textbf{Healthcare App Deployment Model} Mapping of the 3-tier blueprint of our testbed to a healthcare application.}
    \label{fig:HeathAppArch}
\end{figure}

In defining the testbed for the analysis of multi-cloud application deployment, we selected a generic healthcare application because of the technological advancement, breadth of information access, and impact of a potential breach. The healthcare industry has progressed in digitization over the last 20 years, driven in part by White House Executive Order (EO) 13335, \textit{ Incentives for the Use of Health Information Technology}, issued by President G.W. Bush in 2004~\citep{gov13335}. The EO was issued with the motivation to improve the ``quality and efficiency of health care.'' Office management systems and electronic health records are able to store medical procedures and payments. The EO drove the development and industry adoption of these digital systems by offering financial incentives to public and private organizations.

Information security standards in the healthcare industry have been defined in the Security Rule of the Health Insurance Portability and Accountability Act of 1996 (HIPAA). This rule applies to covered entities (CE), who are providers of healthcare services, and business associates (BA) who also have access to the protected health information (PHI) of a CE. CEs and BAs must establish systems, practices, and procedures that ensure the confidentiality, integrity, and availability of PHI. 

The design and architecture of cloud-based systems utilized by CE and BA, which store and process PHI, present a significant challenge to organizations because of the inherently public nature of a cloud deployment~\citep{cloudPHI}. While there is the ability for a cloud provider to create a private subnet for the deployed systems on which the applications run, there is still the fact that the systems are deployed on hardware that is external to the organization and, if not configured properly, could be exposed to public access. While the public nature of the system deployment does not make healthcare applications unique, it does present a significant concern. The nature of PHI makes it some of, if not the most, valued data on the black market. Therefore, the systems that store and process the PHI for organizations must be carefully scrutinized for vulnerabilities. The evaluation of the systems is the core to the risk analysis process.

Figure~\ref{fig:HeathAppArch} identifies the function and application that will run on each cloud provider's system. These applications/functions can run as applications on a full operating system or in a container within a platform system like Docker. As was described in Section~\ref{architecture}, the functionality of the application is spread across multiple cloud providers. Figure~\ref{fig:HeathAppArch} shows the specific patient portal functionality, where the web server and user experience functionality is hosted in an individual cloud provider; the application server, which providers the business logic to the system, runs in a second cloud provider's environment; the database server which manages and issues all data requests runs in a third cloud provider's environment. The patient data repository, where the database server stores the patient data, is in a cloud storage server. This specific implementation of a 3-tier architecture is well-established in cloud applications and includes a presentation tier (user interface), application tier, and data tier~\citep{IBM2023}. This design enables organizations to take advantage of the optimized services by different cloud providers; for example, where one cloud provider would specialize in website hosting and user experience, and another would specialize in database management services. The optimization that is realized through the multi-cloud deployment comes with trades-offs, specifically security risks that have been identified in Table~\ref{tab:risk_analysis}.

In performing a risk analysis, a specific multi-cloud implementation needed to be chosen as the model to be used to assign risk and exposure valuations. The implementation selected was a healthcare patient portal. In real patient portals, users often have the ability to schedule appointments, receive diagnoses and medical test results, as well as pay invoices for care received. The patient portal was chosen because of the breadth of functionality in the patient portal, utilizing personal, financial, and health information of the patient, thus allowing for the risk of a breach to be quantified across many aspects of the architecture. The selection of the patient portal as the testbed allows the evaluation of threat attributes and impact with regard to the testbed as well the overlay of the DREAD framework to model the risk for each treatment with regard to the testbed. In the next section, we describe how the risk analysis is performed and risk scores are calculated.

\subsection{Risk Analysis Methodology}

Utilizing the DREAD Threat Modeling framework, scores were assigned to each attack vector presented in Section~\ref{attackvectors}. For each attack vector, we assigned a score ranging from 0-10 relative to the determined risk presented by that attack vector following the DREAD methodology. To determine the Damage score, we averaged the three damage categories that were assigned and this result was added to the assigned scores for the other DREAD categories to produce the Total Risk score following the formulae, \textit{Total Risk = Average Damage + Reproducibility + Exploitability + Affected Users + Discoverability}. The Total Risk score of the attack vector is then compared to the scores from the other attack vectors to determine the criticality and priority of the mitigation to resolve the threat.

We will first discuss the details of the categorization and damage data used to determine the scores for each of the attack vectors.

\subsubsection{Risk Categories and Damage}\hfill
\label{damage}
The methodology in performing the risk analysis follows a common process of assessing assets (Section~\ref{architecture}) and threats (Section~\ref{attackvectors}). Identifying the intersection of the assets and threats enables the process of determining and quantifying risk and thus enabling the prioritization of these risk and the prescribed mitigations which we have identified in Table~\ref{tab:mitigations}. 

Table~\ref{tab:risk_analysis} documents the Risk Registry. It lists the multi-cloud threats we identified. Table~\ref{tab:stride_cat} shows the  categorizes of these threats within the STRIDE Threat Modeling framework. The STRIDE categorization enables the analysis and impact of each threat with regard to the damage that a threat can incur in the multi-cloud environment. The Total Risk Score for each threat is calculated in Table~\ref{tab:risk_analysis}. The factors that contribute to the score are outlined in the table and in Section~\ref{DREAD}. The risk classification of an attack vector can be critical, high, medium, or low. The damage that can be done by a threat and the probability of the threat make up the total risk score. We will first discuss the attributes of damage and the factors that make up the score.

\subsubsection{Damage Due to Risk}
The damage that results from certain threats can affect different aspects of the operating organization. The legal damage that can occur if a healthcare patient portal is breached can have a significant effect. The healthcare industry has comprehensive security regulations codified in the HIPAA laws. Part of HIPAA is the Security Rule which lays out the security standards, processes, and procedures that a covered entity must follow when handling patient data and managing systems that contain patient data. In the event of data exfiltration, the Office for Civil Rights, the organization responsible for enforcing the HIPAA law, will investigate the incident and determine if the breach is a result of failure to meet the HIPAA Security Rule requirements. Failure to adhere to all the legal requirements when handling/processing/storing patient data can result in compensatory and punitive fines.

The next evaluation of damages is with regard to reputation~\citep{JMIS-reputation}. During an active incident, there is an important role performed by the public relations team. Their objective is to manage the outflow of information with regard to the incident and the response. When this is done poorly, the reputation of an organization can be damaged. When data is stolen from a hospital system, and in our scenario, a breach of the patient portal, there could be a large population of patients who could have their personal, financial, and healthcare information for sale on the dark web. The chance of identity theft is elevated. Individuals who become victims of identity theft because their data was lost by their hospital, will most likely decide to choose another caregiver in the future if given the option. There are actions that an organization can take to preserve their reputation. Focusing on the customer/patient impact is the first step. If the hospital does not help in mitigating the impact on the patient, then their reputation will assuredly drop.

The final damage that is evaluated is to productivity~\citep{Lee2021}. While responding to an incident will take a security analysts away from their daily tasks and developmental projects that could be improving the security of the environment, there is also the aspect of the system slowing down or not working as normal because of the impact of the threat. This can be seen in misconfigurations that cause inter process communication to be interrupted and data not flow across API because of service or data not being delivered at the appropriate time. Damage to productivity will be seen before legal and reputational damage because there is a possibility of interruption of service. While productivity damage is the most visible initially, it does not mean that it will have the largest impact on the organization. If the breach of the system is large enough, then legal and reputational damage could possibly shut down an organization. For a hospital, if the incident is such that the attacker is able to impact a direct patient care system, the impact could be death to a patient in very rare circumstances.

These possible damage scenarios collectively create a total damage impact of an attack. The average of the three damage scores is incorporated into the overall risk score, which will help the enterprise risk manager prioritize the mitigating actions. In the next section, we present the other portions of the DREAD attributes for the analysis and describe how the characteristics of the attack vectors affect their scores.


\begin{table}[h]
  \centering
  \caption{Risk Analysis}
  \label{tab:risk_analysis}
\resizebox{1\columnwidth}{!}{%
\begin{tabular}{lcccccccc}
\multicolumn{1}{l|}{}                                           & \multicolumn{1}{c|}{\multirow{3}{*}{\textbf{Total Risk Score}}} & \multicolumn{3}{c|}{\textbf{Damage}}                    & \multicolumn{4}{c|}{\multirow{2}{*}{\textbf{Threat Attributes}}}                         \\ \cline{3-5}
\multicolumn{1}{l|}{}                                           & \multicolumn{1}{c|}{}                                           & Legal  & Reputation & \multicolumn{1}{c|}{Productivity} & \multicolumn{4}{c|}{}                                                                    \\ \cline{6-9} 
\multicolumn{1}{l|}{\textbf{Description of Threat}}             & \multicolumn{1}{c|}{}                                           & Damage & Damage     & \multicolumn{1}{c|}{Damage}       & Reproducability & Exploitability & Affected Users & \multicolumn{1}{c|}{Discoverability} \\ \hline
Architecture: DoS attacks                                       & 42.67                                                           & 0      & 10         & 10                                & 8               & 8              & 10             & 10                                   \\
Architecture: Differing   Encryption Offerings and Capabilities & 30.33                                                           & 0      & 6          & 7                                 & 7               & 8              & 4              & 7                                    \\
Architecture: CVEs                                              & 44.00                                                           & 0      & 9          & 9                                 & 9               & 10             & 10             & 9                                    \\
Architecture: VPN Infiltration                                  & 25.33                                                           & 0      & 8          & 5                                 & 6               & 9              & 2              & 4                                    \\
Architecture: Guest OS,   Hypervisor, and Host OS               & 22.33                                                           & 0      & 7          & 6                                 & 5               & 8              & 2              & 3                                    \\
Architecture: Addition of   Multiple Cloud Providers            & 19.33                                                           & 0      & 7          & 6                                 & 5               & 6              & 2              & 2                                    \\
API : Interface Format   Consistency                            & 18.00                                                           & 0      & 7          & 8                                 & 2               & 2              & 2              & 7                                    \\
API : Privilege Elevation                                       & 28.00                                                           & 0      & 9          & 6                                 & 8               & 10             & 3              & 2                                    \\
API : Multiple API Connections   Conflict                       & 19.33                                                           & 0      & 5          & 8                                 & 2               & 3              & 2              & 8                                    \\
API : Malformed Packets                                         & 32.00                                                           & 0      & 6          & 9                                 & 8               & 7              & 3              & 9                                    \\
Authentication : Session   Hijacking                            & 23.33                                                           & 0      & 6          & 4                                 & 7               & 8              & 1              & 4                                    \\
Authentication : Substitution   Attack                          & 29.33                                                           & 0      & 7          & 9                                 & 10              & 10             & 2              & 2                                    \\
Authentication :   Man-in-the-Middle                            & 32.67                                                           & 0      & 9          & 5                                 & 7               & 9              & 10             & 2                                    \\
Authentication : Inconsistent   User ACL                        & 24.67                                                           & 0      & 9          & 5                                 & 3               & 9              & 6              & 2                                    \\
Automation : Dynamic changes   to config causing inconsistency  & 27.33                                                           & 0      & 5          & 8                                 & 5               & 8              & 7              & 3                                    \\
Automation : Data poisoning                                     & 34.33                                                           & 0      & 4          & 6                                 & 10              & 10             & 8              & 3                                    \\
Difference in Management:   Service Level Agreement (SLAs)      & 22.67                                                           & 0      & 4          & 4                                 & 4               & 4              & 6              & 6                                    \\
Difference in Management:   Cloud Management Agreement          & 20.67                                                           & 0      & 4          & 4                                 & 4               & 4              & 4              & 6                                    \\
Difference in Management:   Monetization                        & 19.33                                                           & 0      & 5          & 5                                 & 4               & 4              & 4              & 4                                    \\
Difference in Management:   Auto-Scaling                        & 25.67                                                           & 0      & 8          & 9                                 & 6               & 5              & 7              & 2                                    \\
Mismatch in Cyber Legislation:   Data Privacy Laws              & 22.00                                                           & 10     & 6          & 2                                 & 1               & 3              & 6              & 6                                    \\
Mismatch in Cyber Legislation:   Data Control                   & 23.00                                                           & 10     & 6          & 2                                 & 1               & 4              & 6              & 6                                    \\
Mismatch in Cyber Legislation:   Data Release/Sharing           & 23.33                                                           & 10     & 7          & 2                                 & 1               & 4              & 6              & 6                                    \\
Mismatch in Cyber Legislation:   Data Sovereignty Laws          & 22.67                                                           & 10     & 5          & 2                                 & 1               & 4              & 6              & 6                                   
\end{tabular}}
\end{table}


\subsubsection{Risk Due to Attack Profile}\hfill
\label{profile}

The attack profile of a threat incorporates many factors that affect the difficulty of a potential attack. Risk damage, which we discussed in Section~\ref{damage}, directly relates to the attribute of the attack on a vulnerability or, more succinctly, a threat. The different aspects of a threat make up how probable or even possible an attack could happen on a vulnerability. The attribute of the threats we identified for our multi-cloud blueprint is made up of four different factors; Reproducible, Exploitability, Affected Users, and Discoverability. Reproducibility is the ease at which the attack can be reproduced. \textit{Exploitability} is defined as how susceptible the vector is to attack~\citep{DREAD2022}.

\textit{Repeatability} is the attackers are able to exploit the same threat even though it is a known attack vector. The more easily the threat is repeated the higher the score. One reason a threat would score high in this category is if the threat is of a vulnerability that must stay exposed to allow the system to operate normally. Two examples of this happening are when an API does not provide security mechanisms or when patching as a system would cause an application to fail. These vulnerabilities do not have easy or straightforward mitigations and could require rebuilding the system with new software. {Exposure} is the probability of the attacker gaining access to more of the system after the initial breach has occurred. This can be a result of lateral movement in the system, usually caused by credentials being compromised. The higher the possibility of further exposure, the higher the score in this category.

The \textit{Affected Users} attribute is the relative size of the population affected by the attack. The impacted population has a direct impact on the risk score of the attack vector. For example, a Denial of Service (DoS) attack has a higher score for Affected Users than Session Highjacking because a DoS attack can take a website offline, and Session Highjacking only impacts an individual session.

The \textit{Discoverability} of an attack vector is how easy is to discover the targeted vulnerability. A higher score indicates the ease at which valuable information is acquired to support an attack. An example of an easily discovered vulnerability is information found in the public domain or found in the URL of the website. If there is information that could lead to a successful attack, then we would consider that attack vector to have a high score on Discoverability.

Knowing the STRIDE category and the DREAD risk score, we are able to identify mitigations seen in Table~\ref{tab:mitigations} and prioritizations for each of the attack vectors. In the next section, we lay out the technical and administrative mitigation for each of the attack vectors, along with its calculated risk score.

\subsection{Risk Mitigation/Reduction Plan} 
Risk mitigation includes strategies to eliminate the risk, lower the probability of risk occurrence, or lower the impact of the risk on the overall security of the multi-cloud environment. 
Utilizing the information in the risk registry to prioritize the threats, coupled with the specific mitigations for these threats seen in Table~\ref{tab:mitigations}, could allow cybersecurity professionals to develop a mitigation program to reduce the risk from the identified threat. In our healthcare patient portal, there are a number of mitigation techniques that will address the identified threats to the multi-cloud application architecture. In the list below, we have included the risk score and priority. The risk score was determined through the risk analysis performed by the research team. The priority associated with the attack vectors relates to the priority range part of the DREAD Threat Modeling Framework in section~\ref{DREAD}.

In the subsections to follow, we present the threat vectors for each category, the associated risk score calculated in Table ~\ref{tab:risk_analysis}, and its prioritization in relation to where the score lands on the DREAD Threat Modeling Framework. Along with the analysis, we present the counter measures found in the MITRE ATT\&CK Framework for each of the threat vectors. Included with the mitigations are identified technical and administrative counter measures that would reduce the risk related to the threat.

\begin{table}[h]
  \centering
\caption{Attack Vector Countermeasures and Mitigations}
  \label{tab:mitigations}
\resizebox{1\columnwidth}{!}{%
\begin{tabular}{lll}
\textbf{Description of Threat}                                  & \textbf{Countermeasures}                      & \textbf{MITRE ATT\&CK Mitigation}          \\ \hline
Architecture: DoS attacks                                       & WAF w/DDoS mitigation                         & Filter network traffic                     \\
Architecture: Differing   Encryption Offerings and Capabilities & ITIL - Change Management - Secrets Management & N/A                                        \\
Architecture: CVEs                                              & Patch Management - System Hardening           & Patch                                      \\
Architecture: VPN Infiltration                                  & ICAM-MFA, Network segmentation                & Network segmentation, MFA                  \\
Architecture: Guest OS,   Hypervisor, and Host OS               & Patch Management - System Hardening           & User Acct Mgmt                             \\
Architecture: Addition of   Multiple Cloud Providers            & ITIL - Change Management - CMDB               & N/A                                        \\
API : Interface format   consistency                            & ITIL - Change Management - CMDB               & N/A                                        \\
API : Privilege Elevation                                       & PAM - least privilege                         & Monitor, Audit GPO, PAM, User Acct mgmt \\
API : Multiple API Connections   Conflict                       & ITIL - Change Management - CMDB               & N/A                                        \\
API : Malformed Packets                                         & API security \& encryption                    & Monitoring                                 \\
Authentication : Session   Hijacking                            & TLS encryption on all sessions \& MFA         & MFA, delete persistent cookies             \\
Authentication : Substitution   Attack                          & Secure Block-cypher - timestamp               & Audit, PAM, Cert Mgmt                      \\
Authentication :   Man-in-the-Middle                            & Secrets Management - DNSsec                   & Static network config                      \\
Authentication : Inconsistent   User ACL                        & ICAM - SCIM/SAML                              & ICAM                                       \\
Automation : Dynamic changes   to config causing inconsistency  & SOAR Configuration Management - ITIL          & N/A                                        \\
Automation : Data Poisoning                                     & ICAM - Data Encryption - Secrets Management   & Filter network traffic, IPS                \\
Difference in Management:   Service Level Agreement (SLAs)      & ITIL - Service Level Management - CMDB        & N/A                                        \\
Difference in Management:   Cloud Management Agreement          & ITIL - Supplier Management                    & N/A                                        \\
Difference in Management:   Monetization                        & ITIL - Supplier Management                    & N/A                                        \\
Difference in Management:   Auto-Scaling                        & ITIL - Event Management                       & N/A                                        \\
Mismatch in Cyber Legislation:   Data Privacy Laws              & Regulatory Compliance Management              & N/A                                        \\
Mismatch in Cyber Legislation:   Data Control                   & Data Governance                               & N/A                                        \\
Mismatch in Cyber Legislation:   Data Release/Sharing           & Data Governance                               & N/A                                        \\
Mismatch in Cyber Legislation:   Data Sovereignty Laws          & Data Governance                               & N/A                                       
\end{tabular}
}
\end{table}

\begin{table}[htbp]
\tiny
  \centering
  \caption{Threat STRIDE Categorization}
  \label{tab:stride_cat}
\resizebox{0.65\columnwidth}{!}{%
\begin{tabular}{ll}
\multicolumn{1}{l|}{\textbf{Description   of Threat}}           & \textbf{STRIDE Framework Category} \\ \hline
Architecture: DoS attacks                                       & Denial of Service                  \\
Architecture: Differing   Encryption Offerings and Capabilities & Information Disclosure             \\
Architecture: CVEs                                              & ALL                                \\
Architecture: VPN Infiltration                                  & Information Disclosure             \\
Architecture: Guest OS,   Hypervisor, and Host OS               & Tampering with Data                \\
Architecture: Addition of   Multiple Cloud Providers            & ALL                                \\
API : Interface Format   Consistency                            & Tampering with Data                \\
API : Privilege Elevation                                       & Elevation of Privilege             \\
API : Multiple API Connections   Conflict                       & Tampering with Data                \\
API : Malformed packets                                         & Denial of Service                  \\
Authentication : Session   Hijacking                            & Spoofing Identity                  \\
Authentication : Substitution   Attack                          & Denial of Service                  \\
Authentication :   Man-in-the-Middle                            & Information Disclosure             \\
Authentication : Inconsistent   User ACL                        & Elevation of Privilege             \\
Automation : Dynamic changes   to config causing inconsistency  & Denial of Service                  \\
Automation : Data poisoning                                     & Tampering with Data                \\
Difference in Management:   Service Level Agreement (SLAs)      & Repudiation                        \\
Difference in Management:   Cloud Management Agreement          & Repudiation                        \\
Difference in Management:   Monetization                        & Repudiation                        \\
Difference in Management:   Auto-Scaling                        & Denial of Service                  \\
Mismatch in Cyber Legislation:   Data Privacy Laws              & Information Disclosure             \\
Mismatch in Cyber Legislation:   Data Control                   & Information Disclosure             \\
Mismatch in Cyber Legislation:   Data Release/Sharing           & Information Disclosure             \\
Mismatch in Cyber Legislation:   Data Sovereignty Laws          & Information Disclosure            
\end{tabular}
}
\end{table}

\subsubsection{Architecture}
\begin{enumerate}
\item \textbf{DoS Attacks - 42.67 - Critical}. Denial of service attacks seek to overload a system's resources, such as network, CPU, and memory, thereby limiting and possibly preventing a system from responding to service requests. It is difficult to prevent a DoS attack, but the effects of an attack can be reduced and limited through the deployment of technical mechanisms. There are many products and services of Web Application Firewalls (WAF) providing multiple functions, including DoS mitigation. At the core of WAFs is network filtering with specific logic to identify DoS traffic. This filtering can limit the amount of impact that a DoS attack can have on a website.

\item \textbf{Differing Encryption Offerings and Capabilities - 30.33 - High}. Establishing secure connections between systems hosted by different cloud providers can be challenging. The expectation is that the session is secured using the strongest algorithm possible that is common between the two cloud providers. In the event that one cloud provider does not support an algorithm that is not known to have vulnerabilities, then the established session will remain vulnerable or not be established at all. Utilizing change management to maintain strict control over the encryption algorithms that are available within a system and are preferred will help mitigate this threat. Along with change management, secrets management should be employed to support the privacy and control of encryption keys. The capabilities offered by individual cloud providers need to be fully understood when integrating with other cloud providers to enable a cohesive design of the architecture.

\item \textbf{CVEs - 44.00 - Critical}. The common vulnerabilities and exposures (CVE) database holds and scores vulnerabilities and exposures with regard to their impact and exploitability. Vendors respond to these CVEs with remediations through software patches/upgrades or configuration change recommendations. In Table~\ref{tab:risk_analysis}, the Architecture: CVE risk score is the highest of all threats, meaning that this threat is the highest priority to be addressed by remediating the CVEs. The overarching remediation for the system is implementing a patch management process. Because there are new vulnerabilities being found, vendors release patches on a monthly basis, and in the event of critical vulnerabilities, a vendor may make a release off schedule. In many organizations, patch management will have dedicated resources to plan, test, implement, and deploy the necessary patches.

\item \textbf{Virtual Private Network (VPN) Infiltration - 25.33 - High}. VPNs have become the de facto standard for remote secure access. The VPN allows remote users to access organizational assets from outside of the enterprise network, effectively adding the user's device to the enterprise network using a secure encrypted tunnel. Because of the access VPNs give to users, it is a prominent target for attackers. To help mitigate attacks on VPNs, organizations are recommended to implement strong Identity, Credential, and Access Management (ICAM) controls, including MFA. Enabling MFA adds an extra level of authentication on the identity of the person requesting access to the network. If an attacker does breach the VPN, segmenting the network along with enabling access control at the network boundaries prevents attackers from having access to the full enterprise network.

\item \textbf{Guest OS, Hypervisor, and Host OS - 22.33 - Medium}. Cloud providers mainly offer hosting services through virtualization. Virtualization adds layers of hardware abstraction emulated in software. These added layers of software add possible attack vectors as well. IT organizations work to mitigate the possibility of an attacker exploiting these vectors and moving between guest systems within the same host by implementing system hardening. System hardening includes keeping a system updated on patches as well as disabling services that do not need to be run and a strong identity management process.

\item \textbf{Addition of Multiple Cloud Providers - 19.33 - Medium}. Having multiple cloud providers hosting different parts of the same application introduces security challenges. As more cloud providers are added to the environment, the complex the overall security management becomes. Similar to mathematical formulas, the more variables a function has, the more challenging the solution. The mitigation to more cloud providers servicing an application in a multi-cloud environment is to have a higher level of administrative control over the whole environment. Utilizing a Configuration Management Database (CMDB). Utilizing a CMDB, system, and application managers are able to document and monitor the increasing number of system configurations and programmatic interactions with all the other systems helping to mitigate the vulnerabilities introduced.
\end{enumerate}

\subsubsection{API}
\begin{enumerate}
\item \textbf{Interface Format Consistency - 18.00 - Medium}. In the event that there is a difference in the API interface format in what is being sent versus what is expected, the interface will have issues. To reduce the possibility of this happening, a strong change management process should be instituted. The change management process will confirm that the changes to the format are consistent on both sides of the interface, will not lead to data loss, and will protect data integrity. Maintaining a Configuration Management Database (CMDB) should also be used to track how systems are configured and interconnected. The CMDB should contain the history of changes to each interface and the format that is used in these interfaces.

\item \textbf{Privilege Elevation - 28.00 - High}. Part of many breaches is the elevation of a user's privileges. A user's privilege is controlled by their authorizations. Attackers will use this elevated privilege to gain access to data and systems that do not normally have authorizations to access. There are a few mitigations that help to prevent privilege elevation through the manipulation of an API. Deploying a Privilege Access Management (PAM) solution, as well as implementing a least privilege user policy, will help limit a nefarious actor from gaining more access than what they are authorized.

\item \textbf{Multiple API Connections Conflict - 19.33 - Medium}. In Section~\ref{API-Threat}, multiple API connections conflict are described as a vulnerability to be compromised. Having more API connections to manage and maintain leads to overhead for developers. Having multiple API connections can also lead to API confusion, where the naming and construction of the API are not clear~\citep{ACMstaff2016future}. This vulnerability can be mitigated by processes and procedures that we laid out for format consistency, change management, and CMDB.  As we stated in the mitigation for API: Interface format consistency, the use of a CMDB to help document and track the interactions enables the system managers to have full knowledge of all the APIs within the system and their construction and functionality. The information in the CMDB will help mitigate API connection conflicts.

\item \textbf{Malformed Packets - 32.00 - High}. An attacker may inject malformed packets into an interface in an attempt to disrupt services or gain access to the cloud service. To protect the interfaces from malformed packets from an attacker, the interfaces should have security enabled using API key certificates and encryption. The added security will prevent attacker access or disruption to the system. Utilizing certificates and encryption support overall API security and helps prevent system interruption and traffic sniffing.
\end{enumerate}

\subsubsection{Authentication}
\begin{enumerate}
\item \textbf{Session Hijacking - 23.33 - Medium}. Individual session hijack attacks seek to take over a user's session. Attackers use session cookies that have been stolen from a user's computer to impersonate the session/user gaining access to protected data and systems. Mitigations for session hijacking center around the effort to protect session data and user identity. The user of encryption (e.g., SSL/TLS) on sessions as well as Multi-Factor Authentication (MFA) to make sure that nefarious actors are not able to gain or take over user sessions. Besides the operational configuration of encryption and MFA, users should delete persistent session cookies from their system to support the prevention of session hijacking.

\item  \textbf{Substitution Attack - 29.33 - High}. The strength of the cipher algorithm determines whether a substitution attack can be successful. replacing a public key in a message with a new public key that results in an accepted message. This allows an attacker to maliciously send and receive data from a site or service. To protect against this type of attack, an encryption key algorithm should be chosen that does not have very small or impossibility of having a collision. Also, having a timestamp for a message allows the system to determine if the message is an attempt to replay an old message. To prevent these attacks, operational organizations should implement processes like certificate auditing, Privileged Access Management (PAM), and certificate management to maintain control and security of their certificates and encryption keys.

\item  \textbf{Man-in-the-Middle - 32.67 - High}. These attackers are characterized by an attacker silently inserting themselves into a communication between two parties and sniffing the traffic. This is typically accomplished by replacing the Domain Name System (DNS) response to a system to make that system believe that they are interacting with the expected remote system, but in actuality are sending their messages to a system that is intercepting, reading, and relaying the message to the actual remote system without either system realizing that the secure communication has been compromised. A recent innovation has been Domain Names System Security Extensions (DNSSec), where a layer of security is added to Domain Name Service (DNS) to enable and support security around the DNS service. An organization can also enable static network configurations to bypass DNS queries that could have been compromised, thus protecting the communication against this type of attack.

\item  \textbf{Inconsistent User Access Control List (ACL) - 24.67 - High}. Discrepancies between authorization control configurations for user accounts between cloud systems can lead to unauthorized privilege elevation. Mitigation for this type of threat is centralized Identity, Credential, and Access Management (ICAM). Having a central database of access authorization for an identity integrated across all cloud services reduces the risk of an identity having more access than required to perform its assigned functions. Utilizing Security Assertion Markup Language (SAML) and System for Cross-domain Identity Management (SCIM), an Identity Provider (IdP) is able to manage the complete lifecycle of an identity from a central database, maintaining consistent authorization control for users across all integrated systems.
\end{enumerate}

\subsubsection{Automation}
\begin{enumerate}
\item \textbf{Dynamic Changes to Configuration Causing Inconsistency - 27.33 - High}. Organizations are deploying automation solutions to improve the configuration consistency and response time to threats to systems.  Cloud providers deploy tools that manage the deployment of systems. These tools enable IT teams to deploy servers with consistent configuration and, therefore, able to support fluctuations in demand to meet the need of the users. Tools that are available to the IT team can enable dynamic modification of system configuration depending on system performance and user demand. These changes can impact user access and system availability. To mitigate this threat, strict configuration management should be employed. Tools such as Chef, Puppet, Ansible, and Terraform can be used to manage configurations of systems with the environment to make sure that changes are consistent across the systems. Another mitigation procedure would be to document and confirm the cloud configuration of each system and understand the different configuration options of each cloud provider and confirm that the dynamic aspect of the configuration in the cloud provider would not have a detrimental effect resulting in a vulnerability.

\item  \textbf{Data Poisoning - 34.33 - High}. Utilizing automation to manage systems requires the utilization of system and environmental data. An automation process making changes to a system in response to erroneous data could have a deleterious effect on the operation of the application and system. The mitigation of this type of attack requires protection at many different junctures in the environment. Identity protection using ICAM management and MFA, along with data encryption and secrets management, is foundational to preventing a user from manipulating data within the system. Within the network, having an Intrusion Protection System (IPS) along with traffic filtering will provide prevention as well as warning signs of data manipulation that should trigger further investigation and analysis. 

\end{enumerate}

\subsubsection{Difference in Management}
\begin{enumerate}
\item \textbf{Service Level Agreement (SLAs) - 22.67 - Medium}. Cloud providers have SLAs for the services that they provide. These typically include guarantees for uptime, performance, support response time, and credit downtime over the time outlined in the SLA. Managing the differences in the SLA across different cloud providers can be challenging. The difference in uptime between providers can impose a business risk. Depending on the application being hosted, downtime can have a significant financial impact. Mitigating this risk becomes an administrative function. The SLA data from all the providers should be kept in a CMDB so that it can easily be referenced in the event of an outage by the cloud provider. The CMDB supports the Information Technology Infrastructure Library (ITIL) - Service Level Management process as well, enabling organizations to incorporate the management of SLA agreements as part of their full IT administration.

\item  \textbf{Cloud Management Agreement - 20.67 - Medium}. The scope of the Cloud Management Agreement is closely aligned with the Information Technology Infrastructure Library (ITIL) process of Supplier Management~\citep{ITIL}, where the organization's objective is to maintain alignment with business needs, establish relationships, and manage requirements. The risk involved in poor Cloud Management Agreement or Supplier Management is financial as well as operational. Misalignment with the business needs could result in a waste of expenditure on cloud services. Mismanagement of supplier relationships and poorly defined requirements could lead to service interruption or degradation. Mitigation of this risk is to establish a strong supplier management process. Industry best practices for this are codified in the ITIL Supplier Management practice.

\item \textbf{Monetization - 19.33 - Medium}. When first introduced, the initial attraction of cloud-hosted applications was the promised savings an organization could recognize by not having to spend capital funds but move to an operational expense model that could be managed on a granular level that would align more closely to the dynamic aspect of the compute cycle demand. Managing monetization within multiple cloud service providers incurs risks that span business and technical administrative functions. There is a high level of difficulty/risk in implementing consistent control across multiple providers. To mitigate this risk, organizations should implement ITIL Supplier Management. In doing so, the organizations will be able to document and understand the differences in each supplier/provider cost model when the payload expands and shrinks. 

\item \textbf{Auto-Scaling - 25.67 - Medium}. The ability to dynamically increase and decrease the level of resources (CPU, memory, storage, etc.) allocated to an application running in cloud-based systems~\citep{perezsalazar2021dynamic}. Each cloud provider has different configuration and management controls. These controls dictate the dynamic allocation of resources, and as such, each cloud provider has different thresholds and controls for resource allocation. The mitigation recommended for this attack vector is monitoring and event management. By identifying when significant events affect the load and, subsequently, the performance of the application, the cloud application administrators are able to intervene if necessary and bring the disparate cloud environments into balance and allow for the efficient execution of application services. 
\end{enumerate}

\subsubsection{Mismatch in Cyber Legislation}
\begin{enumerate}
\item \textbf{Data Privacy Laws - 22.00 - Medium}. The complexity of Data Privacy Laws continues to increase. In April 2022, there were 18 countries that had specific Data Privacy laws in place~\citep{IAPP2022}. That count did not include any of the EU countries that fall under GDPR. In the United States, there are five states that have data privacy laws: California, Utah, Colorado, Connecticut, and Virginia. The Iowa legislature also passed a bill in 2023 instituting data privacy protections that is waiting for the governor's signature~\citep{Iowa2023}. The risk of violation increases with a constantly changing regulatory landscape as well as the deployment of multiple cloud providers in different principalities. It is critical to have a mature regulatory management program in place to monitor the ever-changing regulations and confirm through internal auditing that the application and systems adhere to the applicable data privacy laws. Failure to meet the requirements of these laws could have a significant financial impact. 

\item \textbf{Data Control - 23.00 - Medium}. Principalities are passing legislation that empowers individuals to dictate how and if an organization is able to store and hold the person's information. These laws also specify that users of the system must be notified what data is being collected by the system and get a user's consent to collect the data. The major risk to an organization is the difference in how these laws across countries and states create challenges in how data is to be managed and maintained internally in the system. To mitigate this risk, a data governance program should be implemented to create policies that will align with the legislation of the concerned principalities. The IT organization is still responsible for following the policies that outline the data control requirements of the laws. 

\item \textbf{Data Release/Sharing - 23.33 - Medium}. Free services such as Google~\citep{google2021} and Facebook~\citep{facebook2018} are in the personal information business. The pervasiveness of these services has driven principalities to address the use of user data. Part of the legislation regarding data and privacy covers how businesses must notify users how collected information will be used by the business and if it will be released or shared with a third party. Organizations' data governance programs must also track and integrate legislation from governing bodies regarding the legality of sharing user data with external third parties. 

\item \textbf{Data Sovereignty Laws - 22.67 - Medium}. Depending on the principality the data is collected, processed, and stored, the laws of a principality could dictate how the data is governed depending on the company's country of incorporation or locality. The risk of being under so many different legal controls is being able to confirm compliance with each of the legal requirements. To mitigate this risk, the Data Governance program of an organization should work to understand the requirements, the overlap, and the priority of the different laws within the principalities that the organization operates. 

\end{enumerate}

\section{Conclusion}
\label{sec:conclusion}

In this paper, we performed a risk and vulnerability analysis of multi-cloud application deployments. After defining and explaining the blueprint for the research, we presented a 3-tier web application architecture as the platform by which we determine, evaluate, and explore the attack vectors. Utilizing the STRIDE threat modeling framework, in Table~\ref{tab:stride_cat} we categorized each attack vector into a STRIDE category enabling the identification of countermeasures and mitigations. We utilized the DREAD threat modeling to perform a risk analysis of the attack vectors, determining the severity of risk associated with the attack vector and prioritizing the remediation of the threats. Applying the STRIDE and DREAD threat modeling methods, we present an analysis of the ecosystem across six attack vectors: cloud architecture, APIs, authentication, automation, management differences, and cybersecurity legislation. Our analysis shows multi-cloud attack vectors that were identified range in severity from medium to critical. The severity can be used to prioritize  remediation. 



\section*{Acknowledgement}
This research was supported by NSA H98230-21-1-0317 through INSuRE+C, and National Science Foundation (NSF) grant \#1565484. The views, opinions, and/or findings expressed are those of the authors and should not be interpreted as representing the official views or policies of the Department of Defense or the U.S. Government.

\bibliographystyle{ACM-Reference-Format}
\bibliography{bibliography.bib}


\end{document}